\documentclass[12pt,a4paper]{article}
\usepackage{epstopdf}
\usepackage[dvips]{graphicx}
\usepackage{amsmath}
\usepackage{epstopdf}
\numberwithin{equation}{section}

\title{Nonlinear analysis of time series  of vibration data from a friction 
brake: SSA, PCA, and MFDFA}
\author{Nikolay K. Vitanov$^{a,b}$, Norbert P. Hoffmann$^{c,d}$, 
Boris Wernitz$^{e}$}
\date{
$^a$ Institute of Mechanics, Bulgarian Academy of Sciences,
Akad. G. Bonchev Str., Bl. 4, 1113 Sofia, Bulgaria \\
$^b$ Max-Planck Institute for the Physics of Complex Systems,
N{\"o}thnitzer Str. 38, 01187 Dresden, Germany \\
$^c$ Department of Mechanical Engineering, Imperial College London, 
London SW7 2AZ, United Kingdom \\
$^d$ Dynamics Group, Hamburg University of Technology,
21073 Hamburg, Germany \\
$^e$ Ferodo Friction,
21073 Glinde, Germany \\
}
\begin{document}
\maketitle
\begin{abstract}
We use the methodology of singular spectrum analysis (SSA), principal 
component analysis (PCA), and  multi-fractal detrended fluctuation 
analysis (MFDFA), for investigating characteristics of vibration time 
series data from a friction brake. SSA and PCA are used to study the 
long time-scale characteristics of the time series. MFDFA is applied 
for investigating all time scales up to the smallest recorded one. 
It turns out that the majority of the long time-scale dynamics, that 
is presumably dominated by the structural dynamics of the brake system, 
is dominated by very few active dimensions only and can well be understood
in terms of low dimensional chaotic attractors. The multi-fractal analysis 
shows that the fast dynamical processes originating in the friction 
interface are in turn truly multi-scale in nature. 
\end{abstract}
\section{Introduction}
Nonlinear effects exist in many natural systems. Because of this
nonlinear science made large advances in the last decades 
\cite{scott}-\cite{kudr90}. And one of most intensively expanding
area of the nonlenear science is connected to the time series analysis.
Singular spectrum analysis (SSA) and principal component analysis (PCA) 
\cite{vaut1} - \cite{vit2} are well known data analysis tools that have 
been successfully applied in research on plasma physics, climate, 
magnetospheric dynamics, microbiology, image analysis, 
industrial process control, etc. \cite{schles} - \cite{russel}. In some areas 
of science and technology, SSA and PCA are still not very popular, 
mostly when there are only small amounts of data with comparatively 
poor quality available. Nevertheless, in the last years these methods 
together with  methods of non-linear time series analysis 
\cite{kantz} and stochastic analysis find increasing number of applications 
for analysis, understanding, and control of complex systems 
(see for an example \cite{boek} - \cite{v2}).
\par
In the present study we analyse vibration data collected from a 
vehicle friction brake under operation. Friction brakes do show 
a rich variety of noise and vibration phenomena \cite{Kinkaid,Chen,HoGa}. 
Especially squeal noise is most unwanted, but also a number of other 
unwanted vibration and noise problems related to different kinds of 
friction affected or friction excited dynamics do exist. 
Industry today employs a multitude of both computer based 
simulation approaches to analyse and improve designs, as well 
as laboratory or testing based techniques. Mostly conventional 
methods both from time, frequency or mixed time-frequency domain are prevailing, like e.g. spectrograms 
or wavelet analysis, see e.g. \cite{diao1,diao2}.
In contrast, only comparatively few studies applying techniques 
from non-linear dynamics and non-linear time-series analysis have been
conducted \cite{Feeny1, Oberst1, Oberst2}, and only the more recent 
studies seem to be based on data obtained from full-scale tests.
\par
The present paper is not focused on the most unwanted noise effects, like squeal or groan, but 
rather on the conditions of normal operation, during which seemingly 
low amplitude random vibrations are generated through the sliding 
motion of brake pads over brake disks. The corresponding acoustic 
signature is usually considered to be largely acceptable from the 
engineering perspective, and only in few occasions the broad band 
noise generated does in fact pose problems to the system's design, e.g. when the resulting noise turns out way too loud and long-lasting after long phases of standstill of the vehicle, very moist weather, or the like \cite{HoGa}.  
Nevertheless, in the present study we focus on this dynamical state 
of the brake system, since it is often considered as the well-behaved 
starting point, from which due to bifurcations the malign states emerge. 
Moreover, it could be expected that the complicated and largely 
unknown small scale and high frequency processes at the sliding 
friction interface leave some dynamical footprints on the system state. 
To better understand the nature and the characteristics of this 
seemingly harmless sliding state is the objective of the present study. 
In spirit we follow an earlier work \cite{Wernitz1}, which suggested 
that the dynamics of steady sliding might be remarkably low-dimensional. 
Although the transition to chaos is well known from studying bifurcation 
behaviour of friction affected systems, see e.g. \cite{Feeny2}, it 
turned out rather surprising that also the irregular dynamics of 
steady sliding might be characterised to a large extent by a model 
with a hand full of degrees of freedom only. To gain further insight, 
in this study we apply further and more recent analysis techniques 
that also allow to shed some light on the possibly multi-scale nature 
of the underlying dynamical processes.  
\section{Measurements and data}
For collecting time series-data of the brake system vibration during 
operation we have used an industrial noise dynamometer. Following our 
previous studies \cite{Wernitz1}, a piezoelectric accelerometer has 
been mounted on the backing plate of a brake pad. After substantial 
efforts with respect to the data acquisition, sampling rates of 200 
kHz have been achieved. This sampling rate is about an order of 
magnitude beyond the limit of audible noise, often used in 
industrial application oriented work. It should safely capture the 
most important parts of the slow low-frequency structural dynamics. 
Moreover, from the present knowledge about the dynamics of frictional 
interfaces in brakes, this sampling rate should also allow at least a 
partial coverage of dynamical processes taking place in the friction 
interface itself. 
\par
Since the experimental set-up and the details of the data 
acquisition have already presented at a number of other places, e.g. \cite{Wernitz1, Wernitz2}, 
we will keep this presentation short here. All data has been collected on a commercial brake test rig with a conventional vehicle friction brake mounted, see Figure 1.

\begin{figure}[t]
\begin{center}
\includegraphics[scale=0.8]{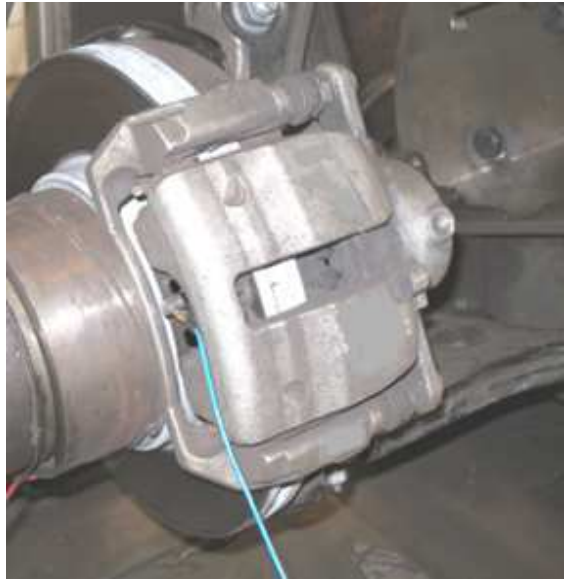}
\end{center}
\caption{Brake system with accelerometer based on the backing plate of the brake pad mounted on commercial brake test rig.}
\end{figure}

The friction noise and vibration data has been obtained by an accelerometer on
the backing plate of the outer brake pad. The sensor is an optimised
piezoelectric type specified with a limiting frequency up to about 100 kHz in
conjunction with a sample rate of the data acquisition above 200 kHz.
The suppression of high frequency electromagnetic radiation effects has been
assured by EMC-compatible countermeasures. In addition, the
galvanic isolation of the chassis earth of the test set-up, the dynamometer itself, 
the data acquisition electronics, the dynamometer automation system as the trigger
source and the isolation of the power electronics demanded further attention and
arrangements. The acceleration signal of the sensor on the brake pad reflects the dynamics of
the backing plate. The sensitivity of the applied sensor strongly decays above 95 kHz, but as the subsequent results suggest, also signal components beyond can be captured. The use of alternative measuring principles, like e.g. optical ones, in order to elevate the cut-off frequency had been accompanied by other problems, like e.g. coherence length effects of laser based approaches. Also the mechanically rough environmental conditions inside the dynamometer test chamber rendered the operation of an optical
laser-doppler-system in the ultrasonic range impossible. As for the position of the sensor, in the following only results that do not depend on the specifics of the sensor location will be reported, and the sensor has been attached right in the middle of the backside of the brake pad's backing plate, as can be seen from Figure 1. The resulting measurements typically yielded large data sets that have not been post-processed in any way, apart from the processing procedures inherent in the signal analysis techniques to be applied. 

As for loading and environmental conditions, three different time-series will be investigated. All of them have been collected from what is called stop-braking in the industry: a constant brake line pressure is applied and the brake disk comes from an initial rotation to a full stop. For all the data, the initial disk rotation rate corresponds to a vehicle speed of 50 km/h. The first time-series, which we will call S3 subsequently, the brake pressure was 30 bar, and the initial disk temperature was 60 degrees Celsius. The second series, S2, has been obtained with 25 bar brake pressure and an initial disk temperature of 100 degrees, the third time series, S1, with 10 bar and an initial temperature of 100 degrees. The three cases have been obtained after typical but different braking test collectives had been run. The motivation for the selected sequence of tests is based on typical braking sequences during normal driving: a first strong brake application is often done with a comparatively cold brake. For the subsequent braking actions, the brake system is heated up due to the previous braking phases, and often softer brake application follows more severe one. Here we would like to mention that of course the present selection of time-series is a bit arbitrary. Many more and quite different loading cases and environmental scenarios would probably be worth while studying. However, the purpose of the present study is rather on showing the feasibility of applying alternative techniques of non-linear time-series analysis to friction vibration data, than on exhaustively analysing brake vibration in general, or even studying a given individual brake configuration completely. Therefore we hope that the results presented subsequently will prove convincing enough to initiate further work in the field. 
  
\par
It has already been shown previously \cite{Wernitz1, Wernitz2} that for each individual braking event the time evolution of the spectral characteristics of the generated vibration is quite stationary. We will thus not repeat the discussion on this behaviour here, but instead directly commence with the novel analysis techniques subject of the present study.

\section{Methods for analysis of studied time series}
First of all we note that if the investigated time series exhibit some 
periodicity, then the autocorrelation function has periodic behaviour too, 
while for chaotic or stochastic time series the autocorrelation function 
usually decays very fast. If long-range correlations are present, we can 
observe a power-law decay. The methods for investigating long-range 
correlations of long enough time series, such as WTMM (Wavelet Transform Modulus
Maxima) method, DFA (Detrended Fluctuation Analysis) and MFDFA (Multi-fractal
Detrended Fluctuation Analysis) have been intensively developed in the last 
decade. For more information on these methods and their applications, see for an
example \cite{kant1} - \cite{yankulova}. Subsequently we will give a very short introduction into the methods selected for the present purpose.
\par
PCA and SSA can be most successfully used to analyse short and non-stationary 
time series in combination with the widely used method of construction of 
phase space by means of delay vectors \cite{k1,k3},\cite{broomhead}-\cite{mees},
often called time-delay phase space construction (TDPSC). The idea of 
SSA-PCA-TDPSC is as follows. Let us have a time series consisting of 
$N^{*}$ values $x(\tau_{0}),x(2\tau_{0}), \dots, x(N^{*} \tau_{0})$ 
recorded by using  fixed time step $\tau_{0}$. On the basis of the time 
series we construct $m-$dimensional vectors as follows. First we choose 
the step $\tau=n \tau_{0}$ and then we construct the vectors ($i=1,2,\dots,n$) 
$$\vec{X}_{i}=\{ x(i \tau_{0}),x(i \tau_{0} + \tau), \dots, x(i \tau_{0} + 
(m-1)\tau)\}$$
By means of the vectors $\vec{X}_i$ we build the trajectory matrix
\begin{equation}\label{tmatrix}
{\bf X} = \frac{1}{N^{1/2}}[\vec{X}_{1}^{T},\vec{X}_{2}^{T},\dots,\vec{X}_{N}^{T}]^{T}
\end{equation}
and the covariance matrix of the trajectory ${\bf K} = {\bf X}^{T} {\bf X}$. 
Let $\vec{k}_{p}$ be the eigenvectors of the covariance matrix and 
$\sigma_{p}$ the eigenvalues corresponding to these vectors. The vectors 
$\vec{k}_{p}$ form an orthonormal basis in the $m-$ dimensional space of 
the vectors $\vec{X}_{i}$. The matrix ${\bf X}$ can be decomposed as: 
${\bf X} = {\bf S} {\bf \Sigma} {\bf C}^{T}$ where ${\bf S}$ is an 
$N \times m$ matrix consisting of the eigenvectors of the trajectory matrix. 
${\rm {\bf C}=[\vec{k}_{1},\vec{k}_{2},\dots,\vec{k}_{m}]}$ is an 
$m \times m$ orthogonal matrix and 
${\bf \Sigma} = {\rm diag[\sigma_{1},\sigma_{2}, \dots, \sigma_{m}]}$ is 
the diagonal matrix constructed by the eigenvalues $\sigma_{i}$, also 
called singular values. These values are non-negative and the common 
rule is to number them in such a way that: 
$\sigma_{1} \ge \sigma_{2} \ge \dots \ge \sigma_{m} \ge 0$.
\par
We can decompose the time series $ \{ x_{i} \}$ using the 
eigenvectors $\vec{k}_{q}$ of the Toeplitz matrix connected to the
time series
\begin{equation}\label{pc1}
x_{i+j}=\sum_{l=1}^{m} a_{i}^{l} k_{j}^{l}, \hskip.5cm 1 \le j \le m
\end{equation}
The principal components $a_{i}^{l}$ of the time series can be obtained
by a projection of the time series on the basis vectors
\begin{equation}\label{pc2}
a_{i}^{l} = \sum_{j=1}^{m} x_{i+j} k_{j}^{l}
\end{equation}
Thus SSA and PCA can be successfully combined with the TDPSC. This can 
be considered the first step in the procedure of time-delay embedding (TDE). 
TDE, however, underlies further restrictions. For an example, the 
dimension of the phase space as well as the time lag for construction 
of delay vectors on the basis of the stationary time series have to be 
chosen by strict procedures \cite{sauer}-\cite{matassini}. For the needs of SSA and PCA the requirements on TDPSC are much looser. One may e.g. choose a small delay (usually $\tau = \tau_{0}$, i.e. $n=1$ ) and large phase space dimension $m$ (as large as the investigated time series allows) \cite{k1}. 
\par
Usually the principal components corresponding to the smaller singular 
values have small amplitude  and oscillate with high frequency. Thus 
the information on large-amplitude slow periodic processes is compressed 
in the first principal components. In many cases, projection on the 
subspace of the first principal components can act as a reasonable filter 
to eliminate the usually low-amplitude and high-frequency processes to be neglected. 
In our case, one might expect the first components to catch the low-frequency structural dynamics, 
while the high-frequency interface processes could be expected to show up 
in the subsequent components. This filtering property of the PCA is very conveniet for the
time-delay embedding methodology for estimation of characteristic quentities of
the system dynamics on the basis of time series. The methodology requires noise-free
time series for a good estimation of corresponding quantities. In many cases the
PCA can be used as a noise filtering tool before application of the time-delay embeding
methods. Such filtering will be used below. We note however that when we come to the application
of the mulrtifractal detrended fluctuation analysis (in Sect. 5) we shall use the original time series
and not the time series processed by the PCA.
\par
In order to perform appropriately the above projection, we have to know the
dimension of the subspace onto which we shall project our time series. 
This dimension is called the statistical dimension \cite{vaut1},\cite{et96}  
and it can be obtained 
in a heuristic fashion on the basis of the singular spectrum analysis (SSA). 
The  singular spectra show us which principal components contain significant 
information for large amplitude and low frequency properties of the time 
series. Usually two situations arise:
\begin{itemize}
\item
Presence of a kink in the singular spectrum.
\par
In many cases in the singular spectra we can observe that there exist several
large values followed by a kink, i.e., the next singular values are much
smaller then the first several ones. The number of large singular values can 
then be taken to determine the statistical dimension $S$ as the dimension
of a principal components phase subspace to which the discussed above 
projection can be performed. Thus, if we observe
a kink in the singular spectrum after a small number of singular values, a 
low-dimensional description of the characteristic features 
of the dynamics that underlies the time series is possible.
\item
There is no kink in the singular spectrum
\par
In this case the selection of the important principal components can be 
based on the requirement that the percentage of the total variance of the 
time series concentrated in the selected number of principal components 
must be larger than some prescribed value, say 90\% or 95\%. We have to 
expect that the dynamics of the system underlying the time series is probably very high-dimensional, i.e., with many significant degrees of freedom.
\end{itemize}
\par
Thus even in the cases where no clear kink is visible in the spectrum of
singular values, we can estimate a value of $S$ which can lead to a 
reasonable subspace projection and to a good model of the dynamics of the 
underlying system.
\par
We note that the statistical dimension provides upper bound for the
minimum degrees of freedom of the measured system \cite{vaut1}. It is
an useful quantity but it is not a characteristic of the system itself
and should not be confused with other dimensions that are independent 
on noise.

\section{SSA and PCA analysis of the data}
\subsection{Influence of the sampling frequency}
\begin{figure}[t]
\begin{center}
\includegraphics[scale=0.8]{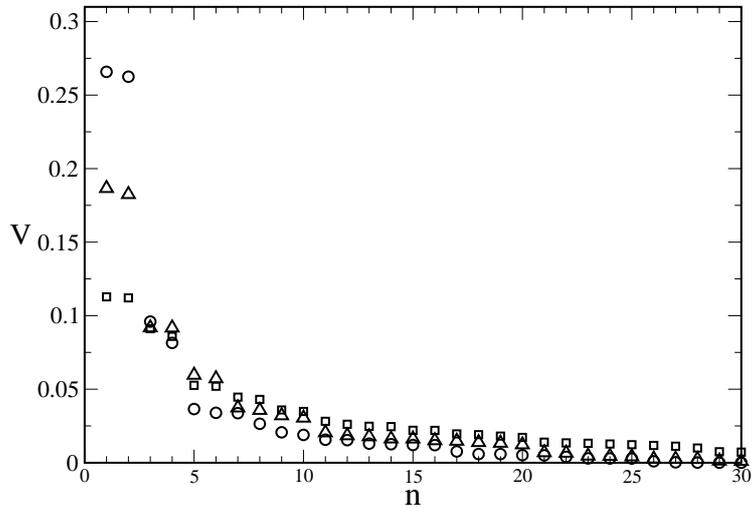}
\end{center}
\caption{Influence of the sampling frequency on the variance connected to the 
principal components. Circles: sampling frequency of the original record.
Triangles: sampling frequency is two times lower in comparison to the
sampling frequency of the original record. Squares: sampling frequency is four
times lower in comparison to the sampling frequency of the original record.}
\end{figure}

Fig. 2 shows the influence of the sampling frequency on the variance connected
to the principal components of the different time series. The figure shows
that the influence of the sampling frequency is quite strong. In principle
one has to expect that the recorded time series will capture processes with
a characteristic frequency up to the sampling
frequency. Visual inspection of the results for the recorded data shows that in the case
of the time series under investigation, the sampling frequency was just enough to capture
all significant macro-scales. If we reduce two times the sampling frequency
we lose substantial information: the variation at the largest principal components drops.
This means that we lose significant informations hidden in the large time scales
of the vibration processes connected with the brake event. If we reduce
the sampling frequency further, the amount of the variance in the largest components
drops further. In turn the amount of variance at the small-amplitude
principal components increases.
\begin{figure}[t]
\begin{center}
\includegraphics[scale=0.8]{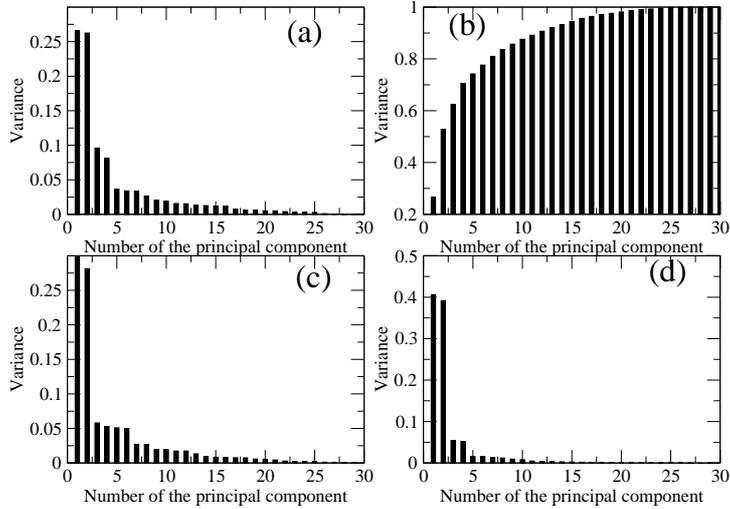}
\end{center}
\caption{Variance connected to the principal components.
Sampling frequency is the same as in the original record. 
Figure (a): time series S1. Figure (b): 
sum of the variances of the principal components for the time series 
S1. Figure (c): time series S2. Figure (d): time series S3.}
\end{figure}
Fig. 3 shows the values of the variance connected to the principal components 
of the investigated time series. As we can see from the Fig. 2b about 90\% of the 
cumulative variance of the time series is contained in the largest 12
principal components. Thus we can use $S=12$ as statistical dimension of the 
time series S1. Similar is the situation with the time series S2. There 
$S=12$ too. The statistical dimension is much smaller for the time series S3, however. For this
time series the $90 \%$ of the variance is contained already in the first 6 principal components, i.e. $S=6$ here. This is an interesting result. One should remember that the time series S3 corresponds to a high pressure brake application in cold conditions, while time series S2 and S1 correspond to brake applications with lower pressures, but with higher initial temperatures. Since the individual braking cases had been extracted from within longer braking tests, it is difficult to pin down the origin of this behaviour further. Nevertheless, the analysis suggests that sometimes higher, and sometimes much lower statistical dimensions may result.
\par
With respect to the primary objective of this study, these are quite remarkable 
findings. First of all the results show that there is a surprisingly small 
number of dominant dimensions, most likely related to the low-frequency 
structural dynamics of the system, capturing most of the relevant dynamics 
of the overall processes, at least in the somehow energetic sense of variance. Second, these low-frequency effects are intricately coupled to the rest of the dynamics, and even the very high sampling rates, which are about a magnitude higher than what simple reasoning based on the limits of the audible range would suggest, do not seem completely sufficient to separate the slow processes of structural dynamics from the fast processes, that most likely are generated in the friction interface itself. Of course, also here it needs to be stressed that the data basis for this study has been comparatively small, with only three time-series at hand. More work with more extended data will thus be needed before further conclusions can be drawn.

\subsection{Embedding after projection}
To further characterise and quantify the data, we use the dominant principal components as base vectors and project the time series on the subspace of the principal components corresponding to largest 
singular values. The information for large-amplitude slow processes is by this compressed 
into the largest principal components. In order to perform appropriately the above projection, 
we have to know the dimension of the subspace onto which we shall project our time series. This dimension is the statistical dimension $S$ calculated above. 
\par
After the projection onto the principal components we can investigate the chaotic characteristics of the time series (filtered from the remaining dynamics by the projection on the largest principal components). In order to do this we shall use the time-delay embedding procedure described below.
Before we start let us note that this procedure doen't lead to one-to-one image of the
corresponding chaotic attractor. The procedudure preserves some topological properties
of the attractor due to the inverainat measures. Some of these measures will be obtained below.

\begin{figure}[t]
\begin{center}
\includegraphics[scale=0.8]{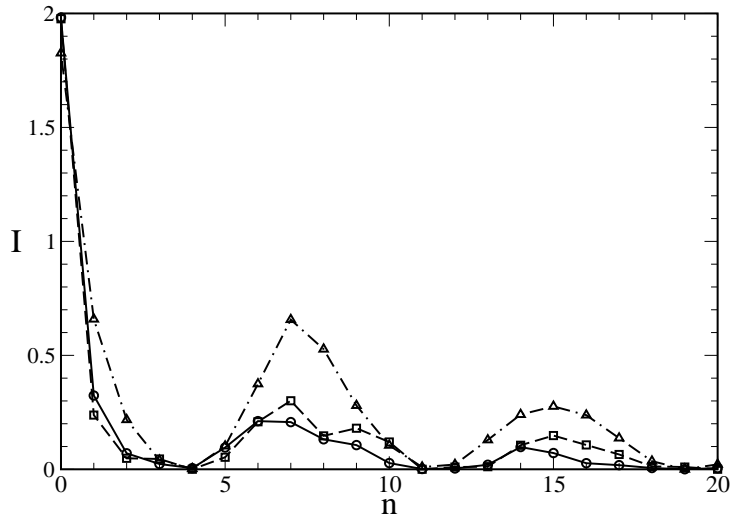}
\end{center}
\caption{The mutual information $I(n)$. Solid line: S1. Dashed line: S2.
Dot-dashed line: S3. }
\end{figure}

In order to perform embedding we need appropriate values of the time-delay and
of the embedding dimension. A candidate value for the time delay can be obtained on
the basis of the quantity called mutual information \cite{fs}
\begin{equation}\label{mi}
M = - \sum_{i,j} p_{ij}(\tau) \ln \frac{p_{ij}(\tau)}{p_i p_j}
\end{equation} 
In order to use Eq.(\ref{mi}) one has to make a partition of the values of the
time series. Then $p_i$ is the probability to find a time series value in the
$i$-th interval of the partition. $p_{ij}$ is the joint probability that if an
observed values falls in the $i$-th interval of the partition then the value
observat at time $\tau$ later falls in the $j$-th interval of the partition.
\par
Figure 4 shows the results for the calculation of the mutual information of 
the investigated projected time series. A good candidate for the delay time in the 
time-delay embedding procedure is obtained from the first minimum of the mutual 
information. One can see that for all time series a good candidate for time delay is $t_{delay}=4$
sampling intervals which corresponds to $2 \cdot 10^{-5}$ s.
\begin{figure}[t]
\begin{center}
\includegraphics[scale=0.8]{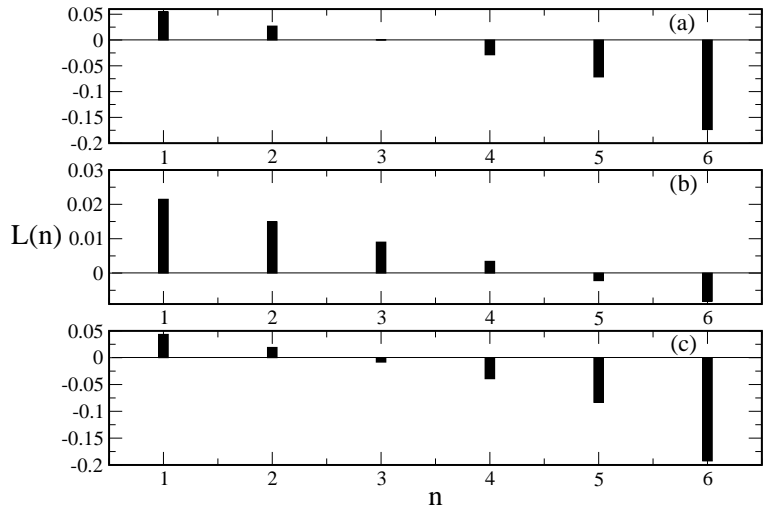}
\end{center}
\caption{The largest 6 Lyapunov exponents for the time series (Sano-Sawada
method). Figure (a): time series S1. Figure (b): time series S2.
Figure (c): time series S3. }
\end{figure}

Next we have to obtain the minimum embedding dimension. For this we shall
use the concept of the false nearest neighbours \cite{kennel}. Let the minimum
embedding dimension for the investigated time series be $m$. In the corresponding $m$-dimensional delay space the reconstructed attractor will be a one-to-one image of the attractor in the original phase space and the topological properties will be preserved. In other words  the neighbours of a given point are mapped onto neighbours in the delay space and neighbourhoods of the points are mapped onto neighbourhoods again. If the embedding is in delay space of dimension $m^* < m$ then the projection the topological structure is no longer preserved. Points are projected into neighbourhoods of other points to which they wouldn't belong in higher dimensions. These points are called false neighbours. Then if we increase the dimension $m^*$ of the delay space the number of the false neighbours will decrease and when we reach the desired dimension $m$ the number of the false neighbours will be very small. Following this algorithm, we have obtained the following values of $m$
for our time series
\begin{itemize}
\item Time series S1: $m=10$
\item Time series S2: $m=12$
\item Time series S3: $m=6$
\end{itemize}

Confirming earlier findings based on different techniques \cite{Wernitz1}, the results again suggest that most of the dynamics is indeed captured within comparatively low-dimensional deterministic dynamics. Although from a purely data focussed perspective this finding might not be too surprising, for the mechanical engineer and a structural dynamics point of view this finding is rather counter-intuitive. The spectral or modal perspective on the vibration properties of a large technical structure like a friction brake would usually feed the expectation that the broad band interface processes would excite a large number of vibration modes of the system. And even within the audible range itself, there would be thousands of modes available to be excited. However, the present findings suggest that the structural dynamics is dominated by a very low-dimensional (sub-)system only.
\par
Additional evidence about the low-dimensionality of the dynamics is obtained by means of 
correlation dimension. Correlation dimension $D_2$ \cite{gp}, \cite{sy93} is a measure of the
dimensionality of the space occupied by a set of random points. For calculation of this dimension
we have used the TISEAN package \cite{k1}. The results for the investigated time series are as follows
\begin{itemize}
\item Time series S1: $D_2 = 3.54 $;
\item Time series S2: $D_2 = 5.43 $;
\item Time series S3: $D_2 = 3.08 $.
\end{itemize}

\subsection{Lyapunov exponents and the Kaplan-Yorke dimension}
The maximum Lyuapunov exponents for the time series have been calculated. The values
are shown in Table 1. Moreover the spectrum of the Lyapunov exponents has been determined on the basis of the methodology from \cite{l1}, \cite{l2}. The results are presented in Fig. 5. We can see that for all time series there are at least two positive Lyapunov exponents. The presence of one positive Lyapunov exponent with the second largest exponent equal to $0$ is already an indicator for the presence of chaos. When more than one positive Lyapunov exponent exists, one sometimes talks about hyper-chaos, which seems to be the case here.
\begin{table}
\begin{tabular} {||c||c|c||}
\hline
\hline
Time series & Maximum Lyapunov exponent & Kaplan-Yorke dimension \\
\hline
\hline
Time series S1 & $L_{max}=0.055$ & $D_{KY}=3.76$  \\
\hline
Time series S2 & $L_{max}=0.021$ & $D_{KY}=5.74$   \\
\hline
Time series S3 & $L_{max}=0.043$ &  $D_{KY}=3.32$  \\
\hline
\hline
\end{tabular}
\caption{Maximum Lyapunov exponents and Kaplan-Yorke dimensions for the 
studied time series.}
\end{table}
\par 
The calculation of the spectrum of Lyapunov exponents allows us also to calculate the
Kaplan-Yorke dimension of the underlying attractors:
\begin{equation}\label{ky}
D_{KY}= k+ \frac{\sum_{i=1}^k L_i}{\mid L_{k+1} \mid}
\end{equation}
where $k$ is the maximum integer such that the sum of the $k$ largest Lyapunov exponents is still non-negative. 
Here one has to deal with the problem of spurious Lyapunov exponents. The problem arises from the 
fact that the rconstructed phase space has extra dimensions compared to the true phase space of the
corresponding physical system. This leads to extra, so called spurious Lyapunov exponents. Recently
Yang, Radons, and Kantz \cite{kr1}, \cite{kr2} have used the concept of covariant Lyapunov vectors
for identification of spurious Lyapunov exponents. The covariant Lyapunov vectors can be calculated
on the basis of algorithm proposed in \cite{gin}. Application of this methodology
leads to the Kaplan-Yorke dimensions for the time series are shown in the third column of
Table 1. What is interesting is that the Kaplan-Yorke dimension jumps for the time series S2 
and has close values for the other two time series. Again, very small numbers result, 
which first of all indicates that the present analysis and projection approach has been successful. 

Again, with respect to the mechanical dynamical system under study, however, the smallness of the attractors dominating the slow structural vibration dynamics is surprising: in engineering dynamics irregular vibration states are usually thought to be caused by truly high-dimensional processes, accessible through statistical rather than deterministic analysis. The vibration response of the structure of a system is generally thought to be such that a large number of degrees of freedom will be involved, when the forcing becomes complex. This way of thinking seems to be originating in the traditional picture of engineering systems to be composed of a large number of undamped and uncoupled oscillators. The present analysis suggests that in fact the resulting dynamics is far less complex in terms of active degrees of freedom than what one could naively think from plainly counting discrete modes in frequency space, or nodes of a geometry capturing finite element representation. While the appearance of such low-dimensional deterministic kernels in complex dynamics seems to have been widely accepted for long time in the sciences, in engineering there is hardly any data based proof for it in technical systems. This is why the present system might be of special interest and importance to the field.    

\begin{figure}[t]
\begin{center}
\includegraphics[scale=0.8]{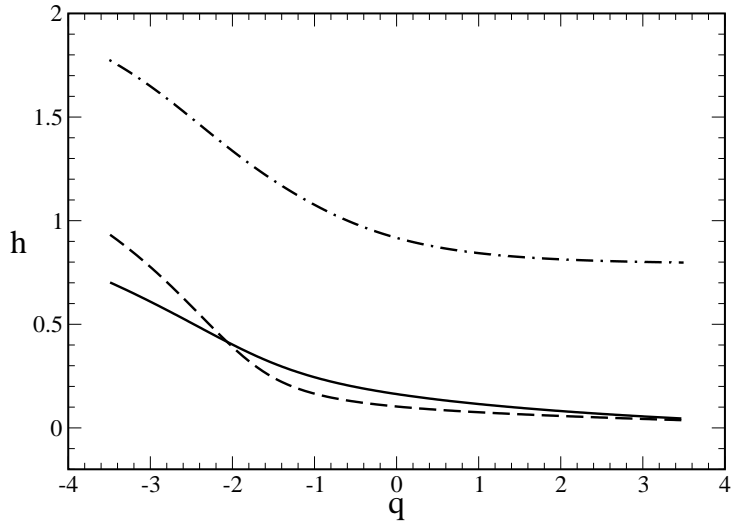}
\end{center}
\caption{$h(q)$ spectra for the time series. Solid line: S1. Dashed line: S2.
Dot-dashed line: S3.}
\end{figure}
\section{Multifractal detrended fluctuation analysis (MFDAFA) of the time series}
\subsection{The MFDFA methodology}
Experience, and also the subsequent results, show that the above approach  
yields interesting insights into the dominant slow time-scale and low frequency dynamics, 
related to the vibrations of the structure. A complementary perspective would be to focus on the dynamical processes taking place at the friction interface itself, and possibly the interactions between these usually multi-scale processes with the comparatively slow structural dynamics. For that purpose, we here employ multi-fractal detrended fluctuation analysis (MFDFA), which has proven highly successful in a number of other scientific disciplines \cite{kant1},
\cite{lopez}, \cite{mur}. The idea here is to investigate the long-range correlations in the inter-maxima intervals in the time series and analyse its correlation properties. 
\par
The long range correlations in time series can be investigated on the basis of Hurst
exponent \cite{bs}, \cite{cran}. When the time series are obtained on the basis of
Brownian motion then the Hurst exponent ie equal to $0.5$. When the process is persistent
the Hurst exponent is larger than $0.5$. When the process is anti-persistent the Hurst
exponent is smaller than 0.5. For white noise the Hurst exponent is 0 and for a simple linear
trend the exponent is 1.
\par
If we have time series with appropriate scaling properties by means of MFDFA we can calculate the spectrum $h(q)$ of the local Hurst exponent \cite{kant1}. Then we obtain the exponent 
$\alpha$ and the fractal spectrum $f(\alpha)$ by means of the relationships
\begin{equation}\label{alpha}
\alpha = h(q) + q \frac{d h}{d q}, \hskip.5cm 
f(\alpha) = q[ \alpha - h(q)] +1
\end{equation}
as well as the mass exponent $\tau_q$
\begin{equation}\label{dq}
\tau_{q} = q h(q) - 1
\end{equation}
The mass exponent supplies evidence for the multifractality of the investigated time series.
The monofractal time series has mass exponent $\tau_q$ with a linear $q$-dependence.
$\tau_q$ with a nonlinear $q$-dependence is evidemce for multifractality.
\par
The realization of the MFDFA method is as follows \cite{kant1} (see also \cite{kant2}-\cite{yankulova}).
First of all, on the basis of the time series $\{x_{k}\}$, we calculate the profile
function
\begin{equation}\label{profile}
Y_{i}= \sum_{k=1}^{i} ( x_{k} - \langle x \rangle), \hskip.5cm i=1,2,\dots, N
\end{equation}

\begin{figure}[t]
\begin{center}
\includegraphics[scale=0.8]{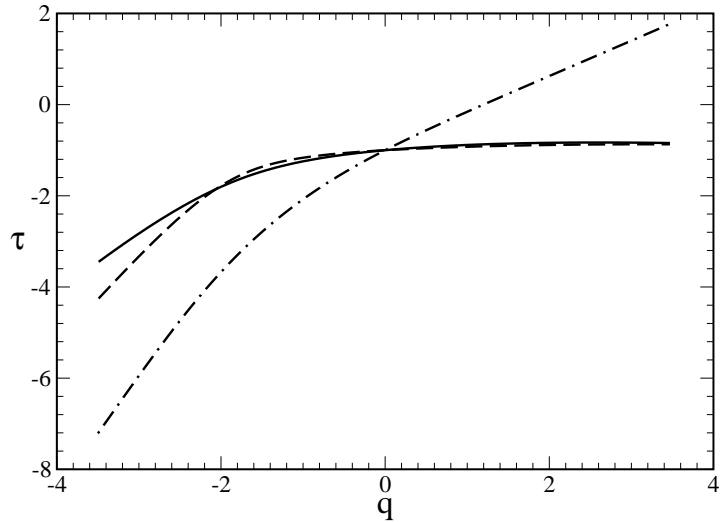}
\end{center}
\caption{$\tau(q)$ spectra for the time series. Solid line: S1. Dashed line: S2.
Dot-dashed line: S3.}
\end{figure}
Then we divide the time series into $N_{s}= {\rm int} (N/s)$ segments and calculate the variation of each segment. As the segments will not include some data at the end of the time series we add additional $N_{s}$ segments, which start from the last value of the time series in the direction of the first value of the series. In order to calculate the variation we have to calculate the local trend (the fitting polynomial $y_{\nu} (i)$) for each segment of length $s$ where $s$ is between some appropriate minimum and maximum values.
The variation is
\begin{equation}\label{var1}
F^{2}(\nu, s) = \frac{1}{s} \sum_{i=1}^{s} \{ Y[(\nu-N)s+i] -
y_{\nu}(i) \}^{2},
\end{equation}
for the first $N_{s}$ segments and 
\begin{equation}\label{var2}
F^{2}(\nu, s) = \frac{1}{s} \sum_{i=1}^{s} \{ Y[N-(\nu-N)s+i] -
y_{\nu}(i) \}^{2},
\end{equation}
for the second $N$ segments. Finally we obtain the $q$-th order fluctuation
function by averaging over all segments as follows
\begin{equation}\label{fl_f}
F_q(s)= \left \{ \frac{1}{2N_s} \sum \limits_{\nu=1}^{2 N_s} \bigg[
F^2(\nu,s) \bigg]^{\dfrac{q}{2}} \right \}^{\dfrac{1}{q}}.
\end{equation}
\par
The scaling properties of $F_{q}(s)$ determine the kind of fractal 
characteristics of the time series. For mono-fractal time series $F_{q}(s)$
scales as $s$ of constant power $h$ for each $q$. For sequences of random
numbers $h=1/2$. If $s >> s^{*}$ this value of $h$ would be unchanged
even in presence of local correlations extending up to a characteristic
range $s^{*}$. If the correlations do not have characteristic length 
$h$ will be different from $1/2$. If the time series exhibit multi-fractal
properties (up to the the smallest value of $s$) the exponent $h$ is not a
constant and becomes a function of the parameter $q$: $h=h(q)$. 
\begin{figure}[t]
\begin{center}
\includegraphics[scale=0.8]{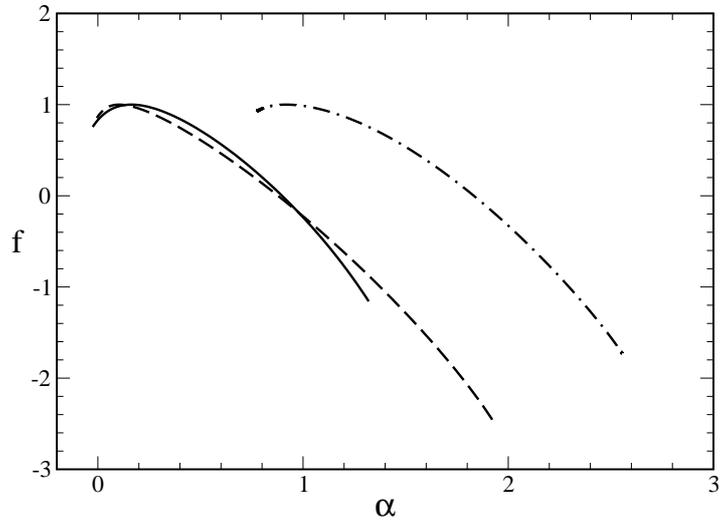}
\end{center}
\caption{$f(\alpha)$ spectra for the time series. Solid line: S1. Dashed line: S2.
Dot-dashed line: S3.}
\end{figure}

\subsection{Results}
Having obtained the above results on the dynamics underlying the slow 
time-scales, we now turn to the multi-fractal analysis of the time series to 
gain further insight into the dynamics from a complementary perspective.
Our goal is just to show that the investigated kind of time series possess
multifractal characteristics. The $h(q)$ spectra of the time series S1, S2, 
and S3 are presented in Fig.6. We observe that $h$ depends on $q$
which means that the time series exhibit multifractal properties up to
smallest time scales recorded. In addition we observe some separation between
the spectra for S3 on one side, and the spectra for time series S1 and S2 from
the other side. This can be seen for an example in Fig. 7 for the $\tau(q)$
spectra, in Fig. 8 for the $f(\alpha)$ spectraq and in Fig. 9 for the
$\alpha(q)$ spectra.
Probably this means that the processes recorded in S1 and S2 are similar on 
a large number of scales, whereas there is something different or additional 
happening for the case of S3. 

\begin{figure}[t]
\begin{center}
\includegraphics[scale=0.8]{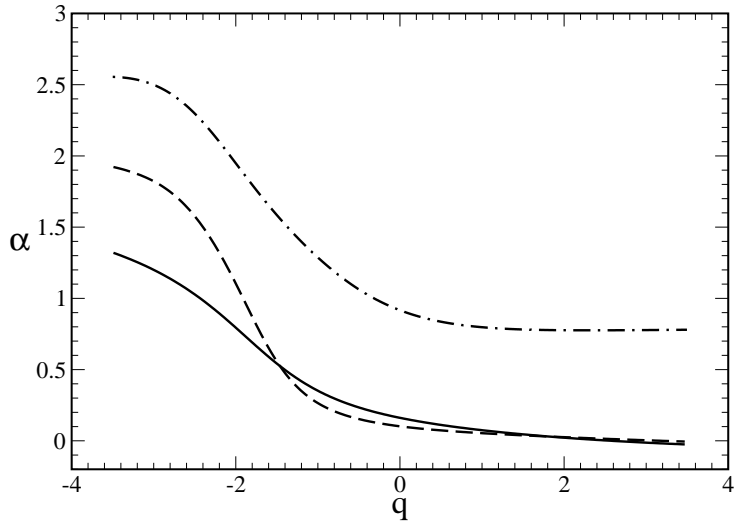}
\end{center}
\caption{$\alpha(q)$ spectra for the time series. Solid line: S1. Dashed line: S2.
Dot-dashed line: S2.}
\end{figure}

\section{Concluding remarks}
In this study we have analysed vibration data of a friction brake under normal operation. Accelerations at a given but arbitrary point on one of the brake pads have been 
measured with a sampling rate of 200 kHz. The recorded time-series have been subjected to a deliberate selection of techniques from non-linear time-series analysis. 
Singular spectrum analysis (SSA) has been combined with principal component analysis (PCA) and time delay phase space construction (TDPSC). Moreover a multi-fractal 
detrended fluctuation analysis (MFDFA) has been conducted. SSA based PCA and TDPSC suggest that the measured data clearly manifests a very small number of 
dominant components, corresponding to the low-frequency structural dynamics. Embedding analysis and calculation of Lyapunov exponents and attractor dimensions 
moreover shows that within the dominant subspace, the dynamics can well be characterised as chaotic, with attractor dimensions well below 10. The higher principal 
components show a rather smooth distribution, indicating the multi-scale nature of the faster processes originating by the sliding dynamics in the friction interface. 
The multi-fractal detrended fluctuation analysis (MFDFA) confirms the multi-scale character of the faster processes.
\par
The results suggest a number of interesting conclusions that need to be discussed. First, although the time-series obtained do seem to represent random processes 
at first sight, it turns out that by the comparatively simple techniques applied the underlying dynamics may easily be separated into a low-dimensional deterministic core 
for the slow, low-frequency part, and a complex multi-scale part. Of course the separation is in no way trivial, and our analysis also shows that the data acquisition should 
indeed be expanded to even higher frequencies to capture an even larger part of the multi-scale processes. Nevertheless, a clear distinction between a small number of 
dominant low-frequency degrees of freedom and separate multi-scale processes can already be drawn from the present analysis. Second, the techniques presented 
might not only be used as tools to allow further insight into the nature of sliding friction, but might also be applied to characterise for example brake pad materials or 
operating conditions. The techniques applied yield a wealth of quantitative results that might well be hoped to be useful in system or load characterisation.
\par
Future work might have a number of directions. First, the measurement basis has to be expanded and even higher sampling frequencies are desirable, also with multi-point measurements. Modelling 
and simulation approaches consistent with the present findings should be developed and model reduction approaches derived. But most importantly we hope that extended approaches along the lines used here will contribute to a better understanding of the today still largely unknown or poorly understood dynamical processes in interfaces, i.e. friction behaviour in general.
\par
Finally let us note that the discussed above results have been obtained by the
help of many computer programs. We have used the software of TISEAN package
\cite{k1} as well as our own software for  multi fractal detrended fluctuation analysis 
of time series.


\begin{thebibliography}{99}
\bibitem{scott}
	A. C. Scott.  Nonlinear  science. Emergence and dynamics of coherent
	structures. Oxford University Press, Oxford, 1999.
\bibitem{murr}
	J. D. Murray. Lectures on nonlinear differential equation models
	in biology. Oxford University Press, Oxford, 1977.
\bibitem{ac}
	M. Ablowitz, P. A. Clarkson. Solitons, nonlinear evolution equations
	and inverse scattering. Cambridge University Press,
	Cambridge, 1991.
\bibitem{vit09a}
	N. K. Vitanov, I. P. Jordanov , Z. I. Dimitrova. On nonlinear population waves.
	Applied Mathematics and Computation. 215, 2009, 2950 -- 2964.		
\bibitem{vx1} 
        N. Martinov, N. Vitanov. On the correspondence between the 
	self-consistent 2D Poisson-Boltzmann structures and the sine-Gordon 
	waves. J. Phys. A: Math. Gen.,  25, 1992 L51 -- L56.
\bibitem{vx2} 
	N. Martinov, N. Vitanov. On some solutions of the two-dimensional 
	sine-Gordon equation. J. Phys. A: Math. Gen.,  25, 1992, L419 -- L426. 
\bibitem{vx3}
	N. Martinov, N. Vitanov. Running-wave solutions of the two-dimensional 
	sine-Gordon equation. J. Phys. A: Math. Gen.,  25, 1992,  3609 -- 3613. 
\bibitem{vx4}
	N. K. Martinov, N. K. Vitanov. New class of running-wave solutions of 
	the 2+1-dimensional sine-Gordon equation. J. Phys. A: Math. Gen., 
	27, 1994,  4611 -- 4618.
\bibitem{vx5}
	N.K. Vitanov. On traveling waves and double-periodic structures in 
	two-dimensional sine-Gordon systems. J. Phys. A: Math. Gen., 1996, 29, 
	5195 -- 5207.
\bibitem{vx6} 
	N.K. Vitanov, N. K. Martinov. On the solitary waves in the sine-Gordon 
	model of the two-dimensional Josephson junction. Zeitschrift f{\"u}r 
	Physik B: Cond. Matt.,100, 1996,129 -- 135.
\bibitem{kudr90}
	N. A. Kudryashov. Exact solutions of the generalized Kuramoto -
	Sivashinsky equation. Phys. Lett. A 147, 1990,  287 -- 291.	




\bibitem{vaut1}
Vautard R., Ghil M. Singular spectrum analysis in nonlinear dynamics with application
	to paleoclimatic time series. Physica D 1989; 35: 395-424.
\bibitem{vaut2}
Vautard R., Yiou P., Ghil M. Singular spectrum analysis - a tookit for short noisy
	chaotic signals. Physica D 1992; 58: 95-126.
\bibitem{vit1}
Vitanov N. K., Sakai K., Dimitrova Z. I. SSA, PCA, TDPSC, ACFA: Useful
	combination of methods for analysis of short and nonstationary time
	series. Chaos Solitons \& Fractals 2008; 37: 187-202
\bibitem{vit2}
Panchev S., Spassova T., Vitanov N. K. Analytical and numerical investigation
	of two Lorenz-like dynamical systems. Chaos Solitons \& Fractals
	2007; 33: 1658-1671.
\bibitem{schles}
Schlesinger M. E., Ramankutty. An oscillation in the global climate system of period
	65-75 years. Nature 1994; 367: 723-726.
\bibitem{krav}
Kravtsov S., Ghil M. Interdecadal variability in a hybrid coupled ocean-atmosphere-sea
\bibitem{marelli}
Marelli L., Bilato R., Franz P., Martin P., Murari A., O'Gorman M. Singular
	spectrum analysis as a tool for plasma fluctuations analysis. Review of
	Scientific Instruments 2001; 72: 499-502.
\bibitem{theobald}
Theobald C. M., Glassey C. A., Horgan G. W., Robinson C. D. Principal component analysis
	of landmarks from reversible images. Journal of the Royal Statistical Society C - 
	Applied Statistics 2004; 53: 163-175.
\bibitem{wold}
Wold S., Esbensen K., Geladi P. Principal component analysis. Chemometrics and 
	Intelligent Laboratory Systems 1987; 2: 37-52.
\bibitem{carland}
Carland J. L., Mills A. L. Classification and characterization of heterothropic
	microbial communities on the basis of patterns of community-level 
	sole-carbon-source utilization. Applied and Environmental Microbiology 1991;
	57, 2351-2359.
\bibitem{russel}
Russell E. L., Chiang L. H., Braatz R. D. Fault detection in industrial processes
	using canonical variate analysis and dynamic principal component analysis.
	Chemometric and Intelligent Laboratory Systems 2000; 51: 81-93.
\bibitem{kantz}
Kantz H., Schreiber T. Nonlinear time series analysis. Cambridge: Cambridge 
	University Press, 1997.
\bibitem{boek}
B\"{o}ck T., Vitanov N. K. Low dimensional chaos in zero-Prandtl-number Benard
	Marangoni convection. Phys. Rev. E 2002; 65: Art. No. 037203.
\bibitem{kantz1}
Kantz H., Holstein D., Ragwitz M., Vitanov N. K. Markov chain model for
	turbulent wind speed data. Physica A 2004; 342: 315-321.
\bibitem{v1}
Dimitrova Z. I., Vitanov N. K. Chaotic pairwise competition.
		Theoretical Population Biology  2004; 66: 1-12.
\bibitem{v2}
Vitanov N. K., Dimitrova, Z. I., Kantz H. On the trap of 
	extinction and its elimination. Phys. Lett. A  2006; 349: 350-355.

\bibitem{Kinkaid}
Kinkaid N. M., O'Reilly O.M., Papadopoulos P. Automotive disc brake squeal. Journal of Sound and Vibration 2003; 267: 105-166.	
	
\bibitem{Chen}
Chen F. Automotive disk brake squeal: an overview. International Journal of Vehicle Design 2009; 51: 39-72.

\bibitem{HoGa}
Hoffmann N., Gaul L. Friction induced vibrations of brakes: research fields and activities. SAE Technical Paper Series 2008; 2008-01-2579: 1-8.

\bibitem{diao1}
Diao K., Zhang L., Meng D. Study on Statistical Analysis of Uncertainty of Disc Brake Squeal. SAE Technical Paper 2014-01-0030.

\bibitem{diao2}
Diao K., Zhang L., Meng D. Modeling and Experimental Investigation on Time-varying Characteristic of Brake Frictional Squeal. SAE Technical Paper 2014-01-2517.

\bibitem{Feeny1}
Feeny B. F, Liang J. W. Phase Space Reconstructions and Stick-Slip. Nonlinear Dynamics 1997; 13: 39-57.

\bibitem{Oberst1}
Oberst S., Lai J. C. S. Chaos in brake squeal noise. Journal of Sound and Vibration 2011; 330: 955-975.

\bibitem{Oberst2}
Oberst S., Lai J. C. S. Statistical analysis of brake squeal noise. Journal of Sound and Vibration 2011; 330: 2978-2994.

\bibitem{Wernitz1}
Wernitz B. A., Hoffmann N. P. Recurrence analysis and phase space reconstruction of irregular vibration in friction brakes: Signatures of chaos in steady sliding. Journal of Sound and Vibration 2012; 331: 3887-3896.

\bibitem{Wernitz2}
Wernitz B. A., Hoffmann N. P. New Approaches to Signal Analysis of Friction Noise and Vibration. Proc. IMechE International Conference of Braking; York 2009.

\bibitem{Feeny2}
Kappagantu R. V, Feeny B. F. Part 1: Dynamical Characterization of a Frictionally Excited Beam. Nonlinear Dynamics 2000; 22: 317-333.
	
	
	
\bibitem{k1}
Hegger R., Kantz H., Schreiber T. Practical implementation of nonlinear time
	series methods: The TISEAN package. CHAOS 1999; 9: 413-435

\bibitem{k3}
Schreiber T. Interdisciplinary applications of nonlinear time series 
methods. Physics Reports 1999; 308: 2-64.

\bibitem{kant1}
Kantelhard J. W., Koscielny-Bunde E., Rego H. H. A., Havlin S., Bunde A.
	Detecting long-range correlations with detrended fluctuation analysis.
	Physica A 2001; 295: 441-454.

\bibitem{kant2}
Kantelhardt J. W., Zschiegner S. A., Koscielny-Bunde E., Havlin S., Bunde A.,
	Stanley HE. Multifractal detrended fluctuation analysis of nonstationary
	time series. Physica A 2002; 316: 87-114.

\bibitem{ph1}
Ivanov P. C., Amaral L. A. N., Goldberger A. L., Havlin S., Rosenblum M. G.,
		Struzik Z. R., Stanley H. E. Multifractality in human heartbeat dynamics.
		Nature 1999; 399: 461-465.

\bibitem{arn1}
Muzy J. F., Bacry E., Arneodo A. The multifractal formalism revisited with
	wavelets. Int. J. Bif. Chaos 1994; 4: 245-302.

\bibitem{yankulova}
	Vitanov N. K., Yankulova E. D. Multifractal analysis of the
long-range correlations of the heartbeat activity of Drosophila 
melanogaster. Chaos Solitons \& Fractals 2006;  28: 768-775.

\bibitem{broomhead}
Broomhead D. S., King G. P. Extracting qualitative dynamics from experimental
	data. Physica D  1986;  20: 217-236.

\bibitem{albano}
Albano A. M., Muench J., Schwant C., Mees A. I., Rapp P. E.
	Singular-value decomposition and the Grassberger-Procaccia algorithm.
	Phys. Rev. A  1988;  38: 3017-3026.

\bibitem{mees}
Mees A. I., Rapp P. E., Jennings L. S. Singular-value
	decomposition and embedding dimension. Phys. Rev A  1987;  36: 340-346.

\bibitem{sauer}
Sauer T., Yorke J. A., Casdagli M. Embedology. J. Stat. Phys. 1991; 65: 579-616.

\bibitem{abarbanel}
Abarbanel H. D. I., Brown R., Sidorowich J. J., Tsirming L. S. The analysis
	of observed chaotic data in physical systems. Rev. Mod. Phys. 1993; 65: 1331-1392.

\bibitem{farmer}
Farmer J. D., Sidorowich J. J. Predicting chaotic time series. Phys. Rev. Lett.
	1987; 59: 845-848.

\bibitem{ks}
Kantz H., Schreiber T. Dimension estimates and physiological data. CHAOS 1995; 5: 143-154.
\bibitem{matassini}
Hegger R., Kantz H., Matassini L., Schreiber T. Coping with nonstationarity by overembedding.
	Phys. Rev. Lett. 2000; 84: 4092-4095.
\bibitem{et96}
Elsner J. B., Tsonis A. A. Singular spectrum analysis. A new tool in time series analysis. New York: Springer, 1996.
\bibitem{fs}
Fraser, A. M., Swinney H. L. Independent coordinates for strange attractors from mutual
        information. Phys. Rev. A 1986; 33: 1134-1140.
\bibitem{kennel}
Kennel M. B., Brown R., Abarabanel H. D. I. determining embedding dimension for phase-space
reconstruction using a geometrical construction. Phys. Rev. A 1992; 45: 3403-3411
\bibitem{gp}
Grassberger P., Procaccia I. Measuring the strangeness of strange attractors. Physica D 1983; 9, 189-208.
\bibitem{sy93}
Sauer T., Yorke J. How many delay coordinates do you need? Int. J. Bifurcation and Chaos 1993; 3, 737-744.
\bibitem{l1}
Sano M., Sawada Y. Measurement of the Lyapunov spectrum from a chaotic time series.
Phys. Rev. Lett. 1985; 55: 1082-1085.
\bibitem{l2}
Eckmann J.-P., Ollifson Kamphorst S., Ruelle D., Ciliberto S. Lyapunov exponents from a
time series. Phys. Rev. A 1986; 34: 4971-4979.
\bibitem{kr1}
Yang H.-L., Radons G., Kantz H. Covariant Lyapunov vectors from reconstructed dynamics:
The geometry behind true and spurious Lyapunov exponents. Phys. Rev. Lett. 2012; 109,
Article No. 244101.
\bibitem{kr2}
Kantz H., Radons G., Yang H.-L. The problem of spurious Lyapunov exponents in time series
analysis and its solution by covariant Lyapunov vectors. J. Phys. A: Math. Theor. 2013;
46, Article No. 254009.
\bibitem{gin}
Ginelli F., Poggi P., Turchi A., Chate H., Livi R., Politi A. Characterizing dynamics
with covariant Lyapunov vectors. Phys. Rev. Lett. 2007; 99, Article No. 130601. 
\bibitem{lopez}
Lopez J. L., Veleva L., Lopez-Sauri D. A. Multifractal detrended analysis of the
corrosion potential fluctuations during copper patina formation on its first stages 
in sea water. International Journal of Electrochemical Science 2014; 9, 1637-1649.
\bibitem{mur}
Murgia J. S., Rosu H. S. Multifractal analysis of row sum signals of elementary cellular 
authomata. Physica A 2012;  391, 3638-3649.
\bibitem{bs}
Bassingthwaighte J. B., Liebovitch L. S., West B. J. Fractal psychology. New York: 
Oxford University Press, 1994.
\bibitem{cran}
Sanchez Cranero M. A., Trinidada Segovai J. E., Garzia Perez J. Some comments on Hurst
exponent and the long memory processes on capital markets. Physica A 2008; 387, 5543-5551.
\end{thebibliography}
\end{document}